\documentclass[onecolumn,showpacs,aps,floatfix,pra,superscriptaddress,nofootinbib]{revtex4}
\usepackage{color}
\usepackage{graphicx}
\usepackage{bm}
\usepackage{amsmath}
\usepackage{enumerate}
\begin{document}
\title{Quantum-classical comparison: arrival times and statistics}
\author{S. V. Mousavi}
\email{vmousavi@qom.ac.ir} 
\affiliation{Department of Physics, The
University of Qom, P. O. Box 37165, Qom, Iran}
\affiliation{ School of Physics, Institute for
Research in Fundamental Sciences (IPM), P.O.Box 19395-5531, Tehran,
Iran } 
\author{S. Miret-Art\'es}
\email{s.miret@iff.csic.es} 
\affiliation{Department of Atomic, Molecular and Cluster Physics, Institute of Mathematics and Fundamental Physics,
Madrid-Spain}
%

%%%%%%%%%%%%%%%%%%%%%%%%%%%%%%%%%%%%%%%%%%%%%%%%%%%%%%%%%%%%%%%%%%%

%
\begin{abstract}
Classical and quantum scattering of a non-Gaussian wave packet by a
rectangular barrier is studied in terms of arrival times to a given
detector location. A classical wave equation, proposed by N. Rosen
[{\it{Am. J. Phys.}} {\bf 32} (1964) 377], is used to study the
corresponding classical dynamics. Mean arrival times are then
computed and compared for different values of initial wave packet
parameters and barrier width. The agreement is improved in the large
mass limit as one expects. A short comment on the possibility of
generalization of Rosen's proposal to a two-body system is given.
Differences in distributions of particles obeying different
statistics are studied by considering a system composed of two free
particles.

\end{abstract}

\pacs{ Schr\"{o}dinger equation, Classical wave equation,
Mean arrival time }

\maketitle

% Section  Section  Section  Section  Section  Section  Section  Section  Section  Section  Section
\section{Introduction}

The classical limit of quantum mechanics is continuously  an
interesting research topic \cite{Ho_book_1993_1, Ba-book-1998,
Home-book-1997}. It is often considered that $\hbar \rightarrow 0$
gives the classical limit. Other limits that are very often used
are large quantum numbers (correspondence principle), short de
Broglie wavelengths and large masses. It has been argued that the
limit $\hbar \rightarrow 0$ is conceptually and mathematically
problematic. In addition, different possible limiting procedures
that can be used in a given problem are mathematically
inequivalent \cite{Ho_book_1993_1}.
%{\bf In Bohmian mechanics, for
%example, the limit of smaller and smaller quantum potentials is
%finally used}.
Very recently, a dividing line between quantum and classical
trajectories in the continuous measurement process has been
proposed leading to the so-called Bohmian time constant
\cite{antonio2013}.

It is well known that the wave function $\psi$ provides statistical
knowledge of the state of a system. The classical analog of this
system corresponds to an ensemble of particles following
deterministic trajectories. Thus, comparison of classical and
quantum mechanics should be meaningful provided that their
statistical predictions for the dynamical evolutions of the same
given initial ensemble is used, apart from some version of the
correspondence principle. Quantum mechanics is formulated in Hilbert
space while classical statistical mechanics is formulated in phase
space. Thus, the corresponding evolution equations are the
Schr\"{o}dinger and Liouville equations, respectively. According to
Feynman in his dynamical theory of the Josephson effect
\cite{FeLeSa-book-1965}, classical and quantum mechanics may be
embedded in the same Hamiltonian formulation by using complex
canonical coordinates \cite{St-RMP-1966, He-PRD-1985}.
Based on Heslot's work \cite{He-PRD-1985}, a theory of
quantum-classical hybrid dynamics has been proposed
\cite{Elz-PRA-2012}, which concerns the direct coupling of classical
and quantum mechanical degrees of freedom. Application of this
proposal to entanglement dynamics and mirror-induced decoherence has
been studied \cite{FrLaEl-arxiv-2014, LaFrEl-arxiv-2014}. 

For a given system, the initial phase space distribution of an
ensemble is not uniquely defined. The simplest choice is to take a
product of the position and momentum distributions. Based on this
scheme, the  quantum-classical correspondence of an arrival time
distribution has been considered in terms of a
non-minimum-uncertainty-product Gaussian wave packet (known as a
squeezed state) which evolves in the presence of a linear potential
\cite{HoPaBa-JPA-2009,Ri-JPA-2013,HoPaBa-JPA-2013}. In particular,
it was shown by Riahi \cite{Ri-JPA-2013} that, if the compared
initial distribution functions are Gaussian with identical
statistical properties, the quantum and the classical mean arrival
times are the same under the influence of at most quadratic
potentials. Moreover, for potentials of the form $V(x) = Ax^2+Bx+C$,
the Liouville equation for the classical phase space distribution
function coincides with the evolution equation for the Wigner
function \cite{Ba-book-1998}. So, if both functions coincide
initially, they will coincide all the time in such a potential. Home
{\it et al.} \cite{HoPaBa-JPA-2009} have taken the initial classical
phase space distribution as the product of the position and the
momentum distributions, and this is the point which has been
criticized by Riahi. This author, in contrast, considers the initial
phase space distribution function to be the Wigner function that is
evaluated using the given initial wave function. In their reply,
Home {\it et al.} state that it is not desirable to use any quantum
input to fix the initial conditions for classical calculations when
one considers classical limit of quantum mechanics.

An alternative route which has been less used in literature, it is
that initially proposed by Rosen \cite{Ro-AJP-1964}. He argues that,
in the large mass limit, the Schr\"{o}dinger equation should be
replaced by another Schr\"{o}dinger-like equation, known as
classical wave equation, which is equivalent to the classical
continuity and Hamilton-Jacobi equations.
He, by speculation, conjectured that
transition from quantum domain to the classical one takes
place for masses of the order of $ m_0 = 2.18 \times 10^{-5}$g or larger.

This classical wave equation contains a non-linear term in $\psi$
which in general prevents superposition of different states unless
these states do not overlap or can be expressed as a multiplication
of each other. Thus, instead of superposition, pure states are
combined to produce mixed states \cite{Ro-AJP-1965}. For instance,
in the scattering of a wave packet by a barrier, if the the energy
is less than the barrier height, the solutions of the classical wave
equation, that is, the incident and reflected functions describe
independent motions, without no interference effects. On the
contrary, in the corresponding quantum problem, the quantum pure
state remains pure and there is no classical limit
\cite{Ho_book_1993_1}.

In this work, our aim is to examine the quantum-classical
correspondence by analyzing the dynamics of a wave packet with
different barrier parameters and masses in terms of mean arrival
times. The detector location is situated well behind the barrier
to prevent some overlapping. We would like to provide a criterion
for the magnitude of mass for which the classical wave equation of
Rosen should be used instead of the Schr\"{o}dinger equation.
Then we consider a two-body system and discuss about the possibility
of generalization of Rosen's classical wave equation. Although
particles are non-interacting, due to the symmetry of the total wave
function spatial correlations exist. By deriving one-body
distributions, we study differences between fermions and bosons.

%===============================================================================
% Section  Section  Section  Section  Section  Section  Section  Section  Section  Section  Section
\section{Scattering of a wave packet by a rectangular barrier and arrival times}

Consider a beam of incident particles  from the left by a
rectangular potential barrier defined as $V(x) = V_0 \Theta(x)
\Theta(a-x)$, where $\Theta(x)$ is the step function and $a$ is the
corresponding width. An ideal detector placed at $x = X~
\gg ~ a$, very far from the barrier, can detect
transmitted particles at a given time. The initial wave function is
assumed to be of the form
\begin{eqnarray} \label{eq: initial_wave}
\psi_0(x) &=& R_0(x) e^{i p_0 x / \hbar} ~,
\end{eqnarray}
in the region $x<0$, which is a plane wave modulated by the
variable amplitude $R_0(x)$ and with initial momentum $p_0$. The
corresponding initial, classical phase space distribution function
is
\begin{equation} \label{eq: initial_phasespace_dis}
  D_0(x, p) = \left \{ \begin{array}{ll}
 |R_0(x)|^2 \delta(p-p_0) , & \mbox{$x < 0$} \\
 0   , & \mbox{$x > 0$.}
 \end{array}
 \right.
\end{equation}

This initial wave function describes {\it classically} a set of
particles having the same momentum $p_0$. {\it Quantum mechanically}
a momentum distribution is usually assumed. When the incident energy
is greater than the barrier height, that is, $E_0 = p_0^2/2m > V_0$,
classically, all particles of mass $m$ ultimately cross the barrier
and arrive at the detector location, $X$. However, quantum
mechanically, the particles can also be reflected by the barrier.
Due to the fact the detector is located very far from the barrier,
then it suffices to consider only the transmitted part of the
initial wave packet.

%===================================================================
% Section  Section  Section  Section  Section  Section  Section  Section  Section  Section  Section
\subsection{Quantum treatment}

Once an incident particle coming from the left of the barrier has
passed completely through it, the transmitted part of the wave
packet is given by \cite{CoDiLa_book_1977}
\begin{equation} \label{eq: QM_transmitted_wave}
\psi_{T}(x, t) = \frac{1}{\sqrt{2\pi}} \int_0^{\infty}
dk~e^{i(kx-E_kt/\hbar)} T(k) \phi(k)~,
\end{equation}
where $E_k = \hbar^2 k^2/2m$, $\phi(k)$ being the Fourier
transform of the initial wave function and
\begin{equation} \label{eq: staionary_trasmission_amplitude}
T(k) = \frac{4 k q e^{i(q-k)a}}{(k+q)^2-(k-q)^2 e^{2iaq}}~,
\end{equation}
is the transmission probability amplitude for monochromatic
incidence with $q = \sqrt{2m(E_k-V_0)/\hbar^2}$.

If the width of the momentum distribution is sufficiently narrow,
the wave packet does not suffer an important distortion or
reshaping \cite{Wi-PR-2006} because of approximate constancy of
the transmission coefficient over the range of the corresponding
integral
\begin{equation} \label{eq: stationary_phase}
\psi_{T}(x, t) = \frac{1}{\sqrt{2\pi}} |T(k_0)| \int_0^{\infty} dk
e^{i[k x - E_k t/\hbar + \eta(k)]} \phi(k)~,
\end{equation}
where $\eta(k)$ is the phase of $T(k)$ and $k_0$ corresponds to
the maximum of $\phi(k)$.

%===============================================================================
% Section  Section  Section  Section  Section  Section  Section  Section  Section  Section  Section
\subsection{Classical treatment}

Rosen \cite{Ro-AJP-1964} discussed and proposed a nonlinear
equation in the configuration space of the form
\begin{equation} \label{eq: cl_wave equation}
i \hbar \frac{\partial \psi}{\partial t} = \left( -\frac{\hbar^2}
{2m} \nabla^2 + V + \frac{\hbar^2}{2m} \frac{\nabla^2
|\psi|}{|\psi|} \right)  \psi~,
\end{equation}
instead of the Schr\"{o}dinger equation for the case of large mass
particles, very often known as the classical Schr\"odinger
equation. If the wave function is written in polar form as $\psi
(x,t)= R(x,t) e^{iS(x,t) / \hbar}$, and is substituted in that
equation one readily obtains
\begin{eqnarray}
\frac{\partial S}{\partial t} + \frac{(\nabla S)^2}{2m} + V &=& 0 ~, \label{eq: cl_HJ1} \\
\frac{\partial \rho}{\partial t} + \nabla \cdot {\bf{j}}  &=& 0,  \label{eq: cl_HJ2}
\end{eqnarray}
which are the classical Hamilton-Jacobi and continuity equations,
respectively. Here $\rho = R^2$ represents the probability density
of an ensemble of trajectories associated with the same
S-function, ${\bf{j}} = \frac{\hbar}{m} \Im(\psi^* \nabla \psi) =
R^2 \frac{\nabla S}{m}$ the probability current density and $S$
the classical action. The real functions $R$ and $S$ are primary
and the classical wave function $\psi$ is deduced from them, i.e.,
it has "a purely descriptive or mathematical significance"
\cite{Ho_book_1993_2}. From Eqs. (\ref{eq: cl_HJ1}) and (\ref{eq:
cl_HJ2}), one easily reaches the continuity of $S$ and
${\bf{j}} \cdot \hat{\bf {n}}$ at a surface where the potential energy
changes discontinuously; $\hat{\bf {n}}$ being the normal to the
surface \cite{Ro-AJP-1965}.

When $E_0 = p_0^2/2m > V_0$, the solution of the classical wave
equation with initial condition (\ref{eq: initial_wave}) is given
by
\begin{equation} \label{eq: classical_wave}
 \psi_{C}(x, t) = \left \{ \begin{array}{lll} R_0\left( x-\frac{p_0 t}{m} \right)
 e^{i[p_0 x  - E_0 t]/\hbar} & \mbox{$x < 0$,} \\
 \sqrt{\frac{p_0}{p^{\prime}_0}} R_0 \left[
\frac{p_0}{p^{\prime}_0}
 \left( x-\frac{p^{\prime}_0 t}{m} \right)  \right]
 e^{i(p^{\prime}_0 x - E_0 t )/\hbar} & \mbox{$0 < x < a$ ,} \\
 R_0 \left[ x-\frac{p_0 t}{m} + a \left( \frac{p_0}{p^{\prime}_0} -
 1\right) \right] e^{i[p_0 x - E_0 t + (p^{\prime}_0-p_0)a]/\hbar} &
\mbox{$a < x$ .}
 \end{array} \right.
\end{equation}
where $p^{\prime}_0 = \sqrt{2m(E_0-V_0)}$ and the continuity of the
action $S$ and current density ${\bf j}$ have been used at the
boundaries $x=0$ and $x=a$. The classical action $S$ generates the
classical trajectory
\begin{equation} \label{eq: classical_trajectory}
 x(t)  = x_0 + u . \left \{ \begin{array}{lll}
 t  ~~~~~~~~~~~~~ , & \mbox{$0 \leq t < -m x_0/p_0$} \\
 \frac{p^{\prime}_0}{p_0}t ~~~~~~~~~ , & \mbox{$-m x_0/p_0 \leq t < m (-x_0/p_0 +a/p^{\prime}_0)$} \\
 t ~~~~~~~~~~~~~~ , & \mbox{$m (-x_0/p_0 +a/p^{\prime}_0) \leq t$ ,}
 \end{array} \right.
\end{equation}
where $ u = p_0/m$ is the velocity of the particle in the free
region. Thus, a classical particle arrives at the detector
location $X$ at time
\begin{equation} \label{eq: tcl_x0}
t_{C}(x_0; X) = \frac{ X-x_0 }{u} + a \left( \frac{1}
{\sqrt{u^2-\frac{2 V_0}{m}}} - \frac{1}{u} \right)~.
\end{equation}
One then readily sees that $t_{C}(x_0; X)$ decreases with the mass
and increases with the barrier width, when other parameters are
kept constant.

%===============================================================================
% Section  Section  Section  Section  Section  Section  Section  Section  Section  Section  Section
\subsection{Arrival times}

In both treatments, the arrival time distribution at the detector
location is given by
\begin{eqnarray} \label{eq: ar_timdis}
\Pi(t; X) &=& \frac{|j(X, t)|}{\int_0^{\infty}dt|j(X, t)|}~,
%=\frac{|j(X, t)|}{|T|^2}
\end{eqnarray}
from which the mean arrival time is calculated,
\begin{eqnarray} \label{eq: mean_ar}
\tau(X) &=& \int_0^{\infty}dt~t~\Pi(t; X)~.
\end{eqnarray}
At this point, it should be noted that in classical mechanics the
concept of arrival time is clear and meaningful. Furthermore, one
can easily {\it prove} that its distribution is given by (\ref{eq:
ar_timdis}). On the contrary, in the standard interpretation of
quantum mechanics, this concept is rather controversial and there
are different {\it proposals} for its definition (see, for
example, Ref. \cite{MuLe-PR-2000} for a review).

Another point that has been noticed in literature is the uniqueness
of the Schr\"{o}dinger probability current density. Demanding that
the non-relativistic current to be the non-relativistic limit of the
unique relativistic current, a unique form has been derived for the
probability current density of spin-$1/2$ \cite{Ho-PRA-1999}; and
spin-$0$ and spin-$1$ particles \cite{StBaNeWe-PLA-2004}. In the
case of spin-$1/2$ particles for a spin eigenstate in the absence of
a magnetic field, the spin-dependent term $(\hbar/m) \Re(\psi^*
\nabla \psi) \times \hat{{\bf{s}}}$ is added to the usual
Schr\"{o}dinger current. Here, $\hat{{\bf{s}}} = \chi^{\dagger}
\hat{\sigma} \chi$ is the spin vector and $\hat{\sigma}$ is the
Pauli matrix. Spin is a quantum-mechanical intrinsic property and
does not have a classical counterpart. Spin-dependent term vanished
in the limit $\hbar \rightarrow 0$ or large mass limit; and in
one-dimensional motion this term does not contribute. So, we put it
away in our calculations.

One can also directly obtain classical mean arrival times without
computing the classical arrival time distribution by means of
\begin{eqnarray} \label{eq: tcl_mean_exact}
\tau_{C}(X) &=& \int dx_0~t_{C}(x_0; X)~[R_0(x_0)]^2 \nonumber \\
&=& \frac{ X- \langle x_0 \rangle }{u} + a \left( \frac{1}
{\sqrt{u^2-\frac{2 V_0}{m}}} - \frac{1}{u} \right)~.
\end{eqnarray}
Just for completeness we mention that the corresponding
fluctuation $\Delta \tau_{C}$ is given by the rms width of the
classical arrival time distribution
\begin{eqnarray} \label{eq: class_fluc}
\Delta \tau_{C} &=& \sqrt{ \int dx_0~[t_{C}(x_0;
X)]^2~[R_0(x_0)]^2
- [\tau_{C}(X)]^2} \nonumber \\
&=& \frac{ \sqrt{ \langle x_0 ^2 \rangle - \langle x_0 \rangle ^2
} }{u} \equiv \frac{\sigma_x}{u} ~,
\end{eqnarray}
which is independent of the detector location and barrier width.

%=================================================================================================
% Section  Section  Section  Section  Section  Section  Section  Section  Section  Section  Section
\subsection{Quantum-classical correspondence for a freely evolving Gaussian packet}
Before going to the general case of scattering of a non-Gaussian
packet by a rectangular barrier, it is instructive at first to
consider free evolution of a Gaussian packet. It's notable that the
results of the previous section are valid here by putting $a=0$ and
$V_0=0$. By solving Eqs. (\ref{eq: cl_HJ1}) and (\ref{eq: cl_HJ2})
one obtains
\begin{eqnarray}
\rho_C(x, t) &=& \rho_0(x-ut) = \frac{1}{\sqrt{2\pi\sigma_0^2}}
\exp{\left[- \frac{(x-x_c-ut)^2}{2\sigma_0^2} \right]} ~, \label{eq: cl_Gauss_free_rho} \\
S_C(x, t) &=& p_0 x - E_0 t~, \label{eq: cl_Gauss_free_S} \\
j_C(x, t) &=& u . \rho_C(x, t) ~, \label{eq: cl_Gauss_free_j}
\end{eqnarray}
for the classical quantities. $\rho_0(x)$ stands for the initial
probability density which is taken to be a Gaussian. Quantum
mechanically one has
\begin{eqnarray}
\rho_Q(x, t) &=& \frac{1}{\sqrt{2\pi\sigma^2}} \exp{\left[-
\frac{(x-x_c-ut)^2}{2\sigma^2} \right]} ~, \label{eq: Q_Gauss_free_rho} \\
S_Q(x, t) &=& S_C(x, t) - \frac{\hbar}{2} \tan^{-1} \left(
\frac{\hbar t}{2m\sigma_0^2}\right) + \frac{(x-x_c-ut)^2 \hbar^2
t}{8m\sigma_0^2 \sigma^2}~, \label{eq: Q_Gauss_free_S} \\
j_Q(x, t) &=& u \left[ 1 + \frac{\hbar^2 t}{4m^2\sigma_0^2 \sigma^2}
\left( \frac{x-x_c}{u} - t \right) \right] . \rho_Q(x, t) ~, \label{eq: Q_Gauss_free_j}
\end{eqnarray}
where $\sigma = \sigma_0 \sqrt{1+ \left( \frac{\hbar
t}{2m\sigma_0^2} \right)^2}$. One clearly sees that in the limit
$\hbar \rightarrow 0$, or in the large mass limit quantum results
approaches to the corresponding classical ones. So, in this specific
example these two limits give the same result.

%=================================================================================================
% Section  Section  Section  Section  Section  Section  Section  Section  Section  Section  Section
\subsection{Quantum-classical correspondence for a non-Gaussian packet}

For practical purposes, we can choose $R_0$ to be a non-Gaussian
function such as \cite{ChHoMaMoMoSi-CQG-2012}
\begin{eqnarray} \label{eq: initial_amplitude}
R_0(x) &=& \frac{1}{(2 \pi \sigma_0^2)^{1/4} \sqrt{1+
\frac{\alpha^2}{2} (1-e^{-\pi^2 /8}) }} \left[ 1 + \alpha \sin
\left( \pi \frac{x-x_c}{4\sigma_0} \right) \right] \nonumber \\
& \times & \exp{ \left[-\frac{(x-x_c)^2}{4\sigma_0^2} \right] }~,
\end{eqnarray}
$\alpha$ being a tunable parameter showing deviation from
Gaussianity; $\sigma_0$ and $x_c$ are the Gaussian ($\alpha = 0$)
wave packet width and center, respectively. This center is chosen
to be  far from the barrier in such a way that there is no
overlapping with the barrier. Our main motivation to use a
non-Gaussian wave packet, apart from being a more general wave
packet, comes from the fact that it is rather difficult to build
exactly Gaussian wave packets in real experiments. This
non-Gaussian function has already been used to study the weak
equivalence principle of gravity in quantum mechanics
\cite{ChHoMaMoMoSi-CQG-2012}. In appendix \ref{sec: app}, some useful
information about this function is provided.

From Eq. (\ref{eq: classical_wave}), the classical wave function
after passing completely through the barrier now takes the form
\begin{eqnarray} \label{eq: cl_wave_nonGaussian}
\psi_{C}(x, t) &=& \frac{1}{(2 \pi \sigma_0^2)^{1/4} \sqrt{1+
\frac{\alpha^2}{2} (1-e^{-\pi^2 /8}) }} \nonumber \\
&\times& \left[ 1 + \alpha \sin \left( \pi \frac{(x-x_c)- u ~ t +
a \left( \frac{p^{\prime}_0}{p_0} -1 \right)}{4\sigma_0} \right)
\right] \nonumber \\
&\times & \exp{ \left\{ i k_0 \left[ x - \frac{u}{2} t + a
\left( \frac{p^{\prime}_0}{p_0} -1 \right) \right] \right\} }
\nonumber \\
&\times & \exp{ \left\{ -\frac{\left[ (x - x_c) - u ~ t + a \left(
\frac{p_0}{p^{\prime}_0} -1 \right) \right]^2}{4\sigma_0^2}
\right\} }~,
\end{eqnarray}
where $k_0=p_0/\hbar$.
In this case, by
using Eqs. (\ref{eq: tcl_mean_exact}) and (\ref{eq: xave_nonG}),
one obtains an analytic relation for the mean arrival time to be
\begin{eqnarray} \label{eq: tcl_mean_exact_nonG}
\tau_{C}(X) &=& \frac{1}{u} \left( X - x_c - \sigma_0 \frac{\pi
e^{3\pi^2/32}\alpha}{\alpha^2(e^{\pi^2/8}-1)+2e^{\pi^2/8}}\right)
\nonumber \\
& + & a \left( \frac{1}{\sqrt{u^2-\frac{2 V_0}{m}}}-1 \right)~.
\nonumber \\
\end{eqnarray}
One sees that $\tau_{C}(X)$ is linear in $X$, $x_c$, $\sigma_0$
and $a$ but not in $\alpha$. This classical mean arrival time decreases
(increases) with $\sigma_0$ for positive (negative) values of
$\alpha$ but there is no dependence on the initial width for a
Gaussian packet ($\alpha=0$). This is approximately true for
a non-Gaussian wave packet with a very large
$\alpha$. In the large mass limit, the classical mean arrival time
reduces to $\tau_{C} = ( X- \langle x_0 \rangle)/ u$ which is
independent of the mass for a given value of $u$. From Eqs.
(\ref{eq: class_fluc}) and (\ref{eq: sigmax}), it is apparent that
the classical fluctuation $\Delta \tau_{C}$ is independent of the
mass, for a given value of $u$, and is minimum for a Gaussian wave
packet.
Using Eqs. (\ref{eq: stationary_phase}) and (\ref{eq:
Fourier_nonGaussian}) one readily obtains,
\begin{eqnarray}  \label{eq: qm_wave_nonGaussian}
\psi_{T}(x, t) &\simeq& \frac{ 1 }{ \sqrt{ 2\pi } } \left(
\frac{2\sigma_0^2}{\pi}  \right)^{1/4}  \frac{ 1 }{ \sqrt{1+
\frac{\alpha^2}{2} (1-e^{-\pi^2 /8}) }} \nonumber \\
&\times& \left\{ f(k_0) - i\frac{\alpha}{2} \left[ f(k_+) - f(k_-)
\right] \right\} ~,
\end{eqnarray}
where
\begin{eqnarray}
f(k_j) &=& |T(k_j)| \sqrt{ \frac{ \pi }{ \sigma_0^2  \left[ 1 +
i\frac{\hbar t }{2m\sigma_0^2} - i \frac{\eta^{\prime
\prime}(k_j)}{2\sigma_0^2}  \right] } } \nonumber \\
& \times& \exp{ \left\{ i k_j \left[x -\frac{\hbar k_j}{2
m}t + \frac{\eta(k_j)}{k_j} \right] \right\} }  ~
\nonumber \\
&\times& \exp{ \left\{ -\frac{[(x-x_c)- \frac{\hbar k_j}{m} t +
\eta^{\prime}(k_j)]^2}{ 4  \sigma_0^2  \left[ 1 + i\frac{\hbar t
}{2m\sigma_0^2} - i \frac{\eta^{\prime \prime}(k_j)}{2\sigma_0^2}
\right] } \right\} }~.
\end{eqnarray}
$\eta^{\prime}(k_j)$ and $\eta^{\prime \prime}(k_j)$ are
respectively the first and second derivative of $\eta(k)$ with
respect to $k$ at $k_j$; $j=0,\pm$. We have used the fact that
according to (\ref{eq: Fourier_nonGaussian}) the amplitude
$\phi(k)$ is a superposition of three Gaussian functions with the
same width $\sigma_k = 1/2\sigma_0$ but with different wave
numbers $k_0$ and $k_{\pm} =k_0 \pm \pi/4\sigma_0$. It has also
been assumed that $\sigma_k$ is sufficiently narrow. Thus, the
corresponding integrals have been extended from $[0, \infty]$ to
$[-\infty, \infty]$ and exponentials Taylor expanded about the
corresponding kick momenta. The reduction of Eqs. (\ref{eq:
cl_wave_nonGaussian}) and (\ref{eq: qm_wave_nonGaussian}) for a
Gaussian wave packet, $\alpha = 0$, shows that:
\begin{itemize}%[i)]
\item The centers of the classical and quantum wave packets differ by an
amount $ a(p_0/p^{\prime}_0 - 1) - \eta^{\prime}(k_0)$.
\item The width of the classical wave packet is constant while the width of
the quantum wave packet increases with time.
\end{itemize}
With these observations classical and quantum packets
, i.e., $|\psi_{C}|^2$ and $|\psi_{T}|^2$ will coincide when
\begin{eqnarray}
 a \left( \frac{p_0}{p^{\prime}_0} - 1 \right) - \eta^{\prime}(k_0)
 &\rightarrow & 0 ~, \\
\left( \frac{\hbar t }{2m\sigma_0^2} - \frac{\eta^{\prime \prime}(k_0)}
{2\sigma_0^2} \right)^2 & \rightarrow & 0 ~,
\end{eqnarray}
and $T(k_0) \simeq 1$.

%=================================================================================================
% SubSection  SubSection  SubSection  SubSection  SubSection  SubSection  SubSection  SubSection  SubSection

\subsection{A two-body non-interacting system}
Schr\"{o}dinger equation for a N-body system in 1D reads
\begin{eqnarray} \label{eq: Sch-twobody}
i\hbar \frac{\partial \Psi(x_1, x_2, \cdots, x_N, t)}{\partial t} &=& \left(
\sum_{i=1}^{N}\frac{-\hbar^2}{2 m_i} \partial_i^2 + V(x_1, x_2, \cdots, x_N, t)
\right) \Psi(x_1, x_2, \cdots, x_N, t)~,
\end{eqnarray}
where $\partial_i = \frac{\partial}{\partial x_i}$. Generalization
of Rosen's proposal to a N-body system is straightforward. By
changing the Schr\"{o}dinger equation as
\begin{eqnarray*} \label{eq: classicalwaveEq-twobody}
i\hbar \frac{\partial \Psi(x_1, x_2, \cdots, x_N, t)}{\partial t} = \left(
\sum_{i=1}^{N}\frac{-\hbar^2}{2 m_i} \frac{\partial^2}{\partial
x_i^2} + V +  \sum_{i=1}^{N}\frac{\hbar^2}{2 m_i}
\frac{\partial_i^2 \Psi} {|\Psi |} \right)
\Psi(x_1, x_2, \cdots, x_N, t)
\end{eqnarray*}
and writing the polar form of the wavefunction $\Psi(x_1, x_2, \cdots, x_N, t) =
R(x_1, x_2, \cdots, x_N, t) e^{i S(x_1, x_2, \cdots, x_N, t)/\hbar}$ one obtains,
\begin{eqnarray} \label{eq: classical HJ-twobody}
\frac{\partial S(x_1, x_2, \cdots, x_N, t)}{\partial t} + \sum_{i=1}^{N}
\frac{[\partial_i S(x_1, x_2, \cdots, x_N, t)]^2}{2m_i} + V(x_1, x_2, \cdots, x_N, t) &=& 0~, \nonumber \\
\frac{\partial R^2(x_1, x_2, \cdots, x_N, t)}{\partial t} +  \sum_{i=1}^{N}
\partial_i \left( R^2(x_1, x_2, \cdots, x_N, t) \frac{\partial_i S(x_1, x_2, \cdots, x_N, t)}{m_i} \right) &=& 0~.
\end{eqnarray}

Now consider a 1D system composed of two free identical particles.
Classically, particles are distinguishable and obey classical
Maxwell-Boltzmann (MB) statistics. Quantum mechanically they are
indistinguishable and obey different statistics. Fermions (Bosons)
obey Fermi-Dirac (Bose-Einstein) statistics for which the total
wavefunction must be antisymmetric (symmetric) under the exchange of
particles in the system. Since particles do not interact, one can
construct solutions of the Schr\"{o}dinger equation from two
single-particle wavefunctions $\psi_a$ and $\psi_b$ as follows:
\begin{eqnarray}
\Psi_{\mathrm{MB}}(x_1, x_2, t) &=& \psi_a(x_1, t) \psi_b(x_2, t)~,\\
\Psi_{\pm}(x_1, x_2, t) &=& N_{\pm}[\psi_a(x_1, t) \psi_b(x_2, t) \pm \psi_b(x_1, t) \psi_a(x_2, t)]~,
\end{eqnarray}
where upper (lower) sign stands for BE (FD) statistics. If
single-particle wavefunctions $\psi_a(x_1, t)$ and $\psi_b(x_2, t)$
are solutions of classical one-body wave equation (\ref{eq: cl_wave
equation}) then $\psi_a(x_1, t) \psi_b(x_2, t)$ and $\psi_b(x_1, t)
\psi_a(x_2, t)$ separately satisfy eq. (\ref{eq:
classicalwaveEq-twobody}), but due to the non-linearity of this
equation symmetrized wavefunctions $ \Psi_{\pm}(x_1, x_2, t) $ are
not solutions of classical wave equation, i.e., there is no
corresponding classical wave equation for indistinguishable
particles as one expects.

Due to indistinguishability of identical particles for the quantum
BE and FD statistics, one-body density (density probability for
observing a particle at point $x$ irrespective of the position of
the other particle) is given by \cite{Zelev-book-2011}
\begin{eqnarray} \label{eq: rho_twobody}
\rho_1(x, t) &=& \int dx_1 dx_2~\delta(x-x_1)~|\Psi(x_1, x_2, t)|^2
= \frac{1}{2} \left[ \int dx_2|\Psi(x, x_2, t)|^2 + \int
dx_1|\Psi(x_1, x, t)|^2 \right]
\nonumber \\
&=& |N_{\pm}|^2 \left(~|\psi_a(x, t)|^2 + |\psi_b(x, t)|^2 \pm
2~\Re\left[ \langle \psi_a(t) | \psi_b(t) \rangle ~ \psi^*_a(x, t)
\psi_b(x, t) \right]~\right) ~,
\end{eqnarray}
where $\langle \psi_a(t) | \psi_b(t) \rangle = \int dx ~ \psi^*_a(x,
t)  \psi_b(x, t) $. It must be noted that probability distributions
for finding a particle at a point $x$ are different for two
particles in the classical MB statistics,
\begin{eqnarray}
\rho_1^{(1)}(x, t) = |\psi_a(x, t)|^2~,
~~~~~~~~~~~~~~~~~~\rho_1^{(2)}(x, t) = |\psi_b(x, t)|^2~.
\end{eqnarray}
By straightforward algebra one gets a continuity equation for
one-body density,
\begin{eqnarray}
\frac{\partial \rho_1(x, t)}{\partial t} + \frac{\partial}{\partial
x} j_1(x, t) &=& 0~,
\end{eqnarray}
where,
\begin{eqnarray} \label{eq: prob_qm_twobody}
j_1(x, t) &=& \frac{\hbar}{2m} \Im \left \{ \int dx_2 \Psi^*(x, x_2,
t) \frac{\partial \Psi(x, x_2, t)}{\partial x} + \int dx_1
\Psi^*(x_1, x, t) \frac{\partial \Psi(x_1, x, t)}{\partial x} \right
\}
\nonumber \\
&=& \frac{\hbar}{m} |N_{\pm}|^2 \Im \left \{ \psi^*_a \frac{\partial
\psi_a}{\partial x} + \psi^*_b \frac{\partial \psi_b}{\partial x}
\pm \langle \psi_a(t) | \psi_b(t) \rangle ~ \psi^*_b \frac{\partial
\psi_a}{\partial x} \pm \langle \psi_b(t) | \psi_a(t) \rangle ~
\psi^*_a \frac{\partial \psi_b}{\partial x} \right \}~,
\end{eqnarray}
and we have used the fact that the wavefunction becomes zero as $x
\rightarrow \pm \infty$. For distinguishable particles obeying
classical MB statistics one has two probability current densities
for each particle,
\begin{eqnarray}
j_1^{(1)}(x, t) = \frac{\hbar}{m} \Im \left \{ \psi^*_a(x, t)
\frac{\partial \psi_a(x, t)}{\partial x} \right \} ~, ~~~~~~~~~~~
j_1^{(2)}(x, t) = \frac{\hbar}{m} \Im \left \{ \psi^*_b(x, t)
\frac{\partial \psi_b(x, t)}{\partial x} \right \}~,
\end{eqnarray}
Noting eqs. (\ref{eq: rho_twobody}) and (\ref{eq: prob_qm_twobody}),
one sees that as long as the single-particle wavefunctions has
negligible overlap, i.e., $ \langle \psi_a(t) | \psi_b(t) \rangle
\simeq 0$ then there is no need for symmetrization. As a result one
can ignore distinguishability of particles and thus uses MB
statistics for which motions of particles are independent and
everything is the same as the one-body systems which was discussed
in previous sections. Thus, we will not consider such a statistics
anymore.

By taking the one-particle wavefunctions as Gaussian packets,
\begin{eqnarray} \label{eq: Gauss_packet}
\psi_i(x, t) = (2\pi s_{ti}^2)^{-1/4} \exp{ \left[ {i k_i(x-u_i
t/2)-\frac{(x-u_i t - x_{ci})^2}{4s_{ti}\sigma_{0i}}} \right] },~~~~
s_{ti} = \sigma_{0i}(1+i\hbar t/2m\sigma_{0i}^2)~,
\end{eqnarray}
with $i={a, b}$, the overlap integral is given by,
\begin{eqnarray} \label{eq: innner-product-psia-psib}
\langle \psi_a(t) | \psi_b(t) \rangle &=& \sqrt{ \frac{2 \sigma_{0a}
\sigma_{0b} }{ \sigma_{0a}^2 + \sigma_{0b}^2 } }
~\exp{\left[-\frac{4(k_a-k_b)^2\sigma_{0a}^2 \sigma_{0b}^2 +
(x_{ca}-x_{cb})^2 + 4i(k_a-k_b)(\sigma_{0b}^2 x_{ca} + \sigma_{0a}^2
x_{cb})}{4(\sigma_{0a}^2 + \sigma_{0b}^2)} \right]} ~.
\end{eqnarray}
It is seen that overlap integral $\langle \psi_a(t) | \psi_b(t)
\rangle$ is independent of time and normalization constants are
given by
\begin{eqnarray} \label{eq: normalization-constants}
N_{\pm} &=& \frac{1}{\sqrt{2}}\left\{ 1 \pm \frac{2\sigma_{0a}
\sigma_{0b}}{\sigma_{0a}^2 + \sigma_{0b}^2}
\exp{\left[-\frac{4(k_a-k_b)^2\sigma_{0a}^2 \sigma_{0b}^2 +
(x_{ca}-x_{cb})^2}{2(\sigma_{0a}^2 + \sigma_{0b}^2)} \right]}
\right\}^{-1/2}~.
\end{eqnarray}
%

%=================================================================================================
% Section  Section  Section  Section  Section  Section  Section  Section  Section  Section  Section
\section{Results and discussion}

For the calculations, the following parameters are kept fixed:
$x_c = -50$ $\AA$, $V_0 = 5$ eV, $u = 4.52 \times 10^{-3} c$ and
$X = 75$ $\AA$,
where $c$ is the light velocity in vacuum. Mean arrival time is
computed quantum mechanically and classically for different masses
and barrier widths. Results are expressed by using the following
units: time is given in femtosecond, length in Angstrom and mass
in MeV$/$c$^2$.

%**********************************************************
\begin{figure}
 \begin{center}
 \includegraphics[width=10cm,angle=-90]{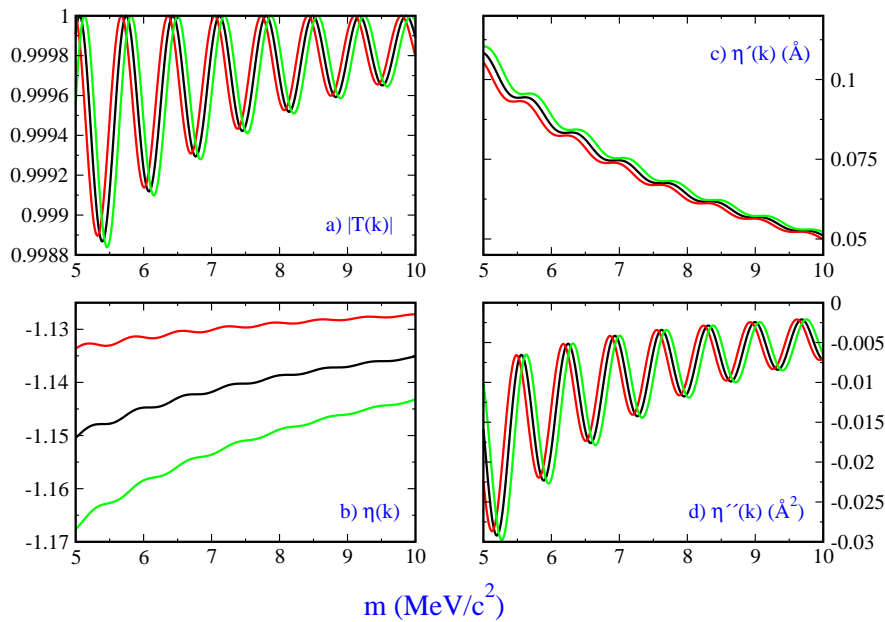}
 \caption{\label{fig1}
%\begin{figure}
%\centering
%\includegraphics[width=12cm,angle=-90]{Figure1.eps}
(Color online) a) $|T(k)|$ b) $\eta(k)$ , c) $\eta^{\prime} (k)$
and d) $\eta^{\prime \prime}(k)$ for $k=k_0$ (black curve),
$k=k_+$ (red curve) and $k=k_-$ (green curve) versus mass for $a =
2${\AA}.} \vspace*{0.7cm} \label{fig: transprob}
\end{center}
\end{figure}

%**********************************************************

We first present some information about the transmission
probability for the wave numbers $k_0$ and $k=k_{\pm}$. As Eq.
(\ref{eq: qm_wave_nonGaussian}) shows, the probability density and
probability current density depend on the modulus of transmission
amplitude and its phase; and the first and the second derivative
of the phase at $k=k_0$ and $k=k_{\pm}$. Figure \ref{fig: transprob}
shows theses quantities as a function of the mass for a given
value of barrier width. One sees all of these parameters oscillate
as mass changes and oscillations become small as mass increases.
At this point it is worth mentioning that the transmitted wave
packet looks like the one for free propagation with minor
differences in between their extrema (compare Eqs. (\ref{eq:
qm_wave_nonGaussian}) and (\ref{eq: free_nonGaussian})).
% and note figure \ref{fig: Fig1}.

%**********************************************************

%**********************************************************
% Figure 2
\begin{figure}
\begin{center}
\includegraphics[width=10cm,angle=-90]{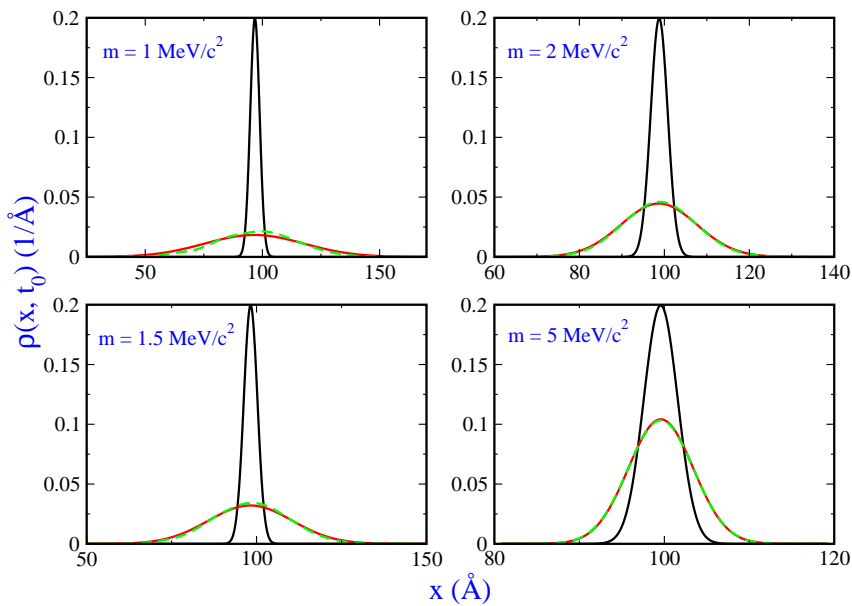}
\caption{(Color online) Probability density for a Gaussian wave
packet versus $x$ at time $t_0 = 11.07$ fs for $\sigma_0 = 2${\AA}
and $a = 8${\AA}; and for different values of mass. Black curve is
the classical wave (\ref{eq: cl_wave_nonGaussian}), red curve is
the quantum wave computed by (\ref{eq: qm_wave_nonGaussian}) and
the green one computed by (\ref{eq: QM_transmitted_wave}). One
sees that red and green curves coincide as mass increases to a
value $m \geq 5$ MeV$/$c$^2$.} \vspace*{0.7cm} \label{fig: rho_x}
\end{center}
\end{figure}

%**********************************************************
% Figure 3
\begin{figure}
\begin{center}
\includegraphics[width=10cm,angle=-90]{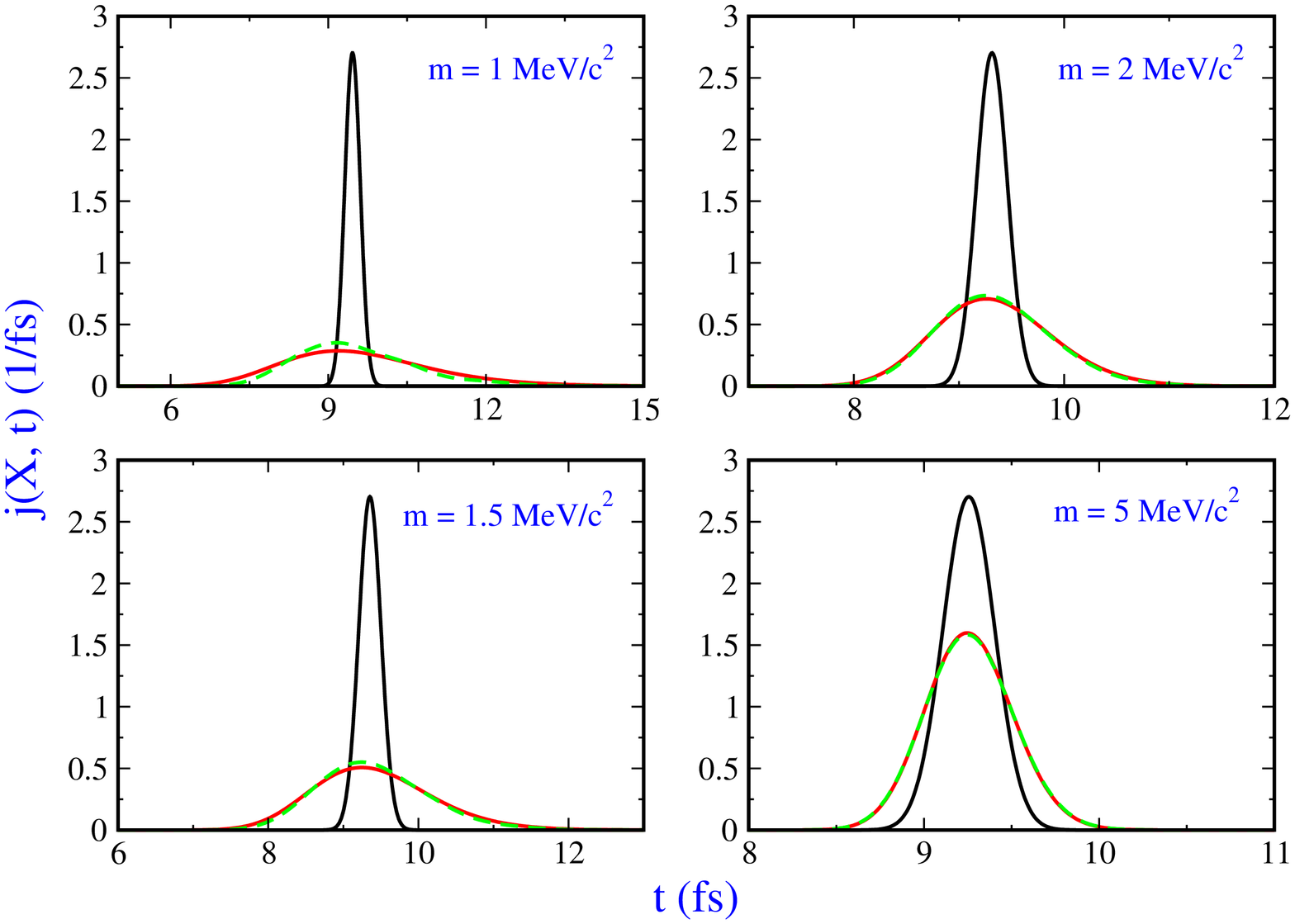}
\caption{(Color online) Probability current density for a Gaussian
wave packet versus $t$ at detector location $X = 75${\AA} for
$\sigma_0 = 2${\AA} and $a = 8${\AA}; and for different values of
mass. Black curve is the classical result while the red one is the
quantum result computed by the wave function (\ref{eq:
qm_wave_nonGaussian}) and the green curve shows the quantum result
computed by the wave function (\ref{eq: QM_transmitted_wave}).}
\label{fig: cur_t}
\end{center}
\end{figure}
%**********************************************************

\begin{figure}
\begin{center}
\includegraphics[width=10cm,angle=-90]{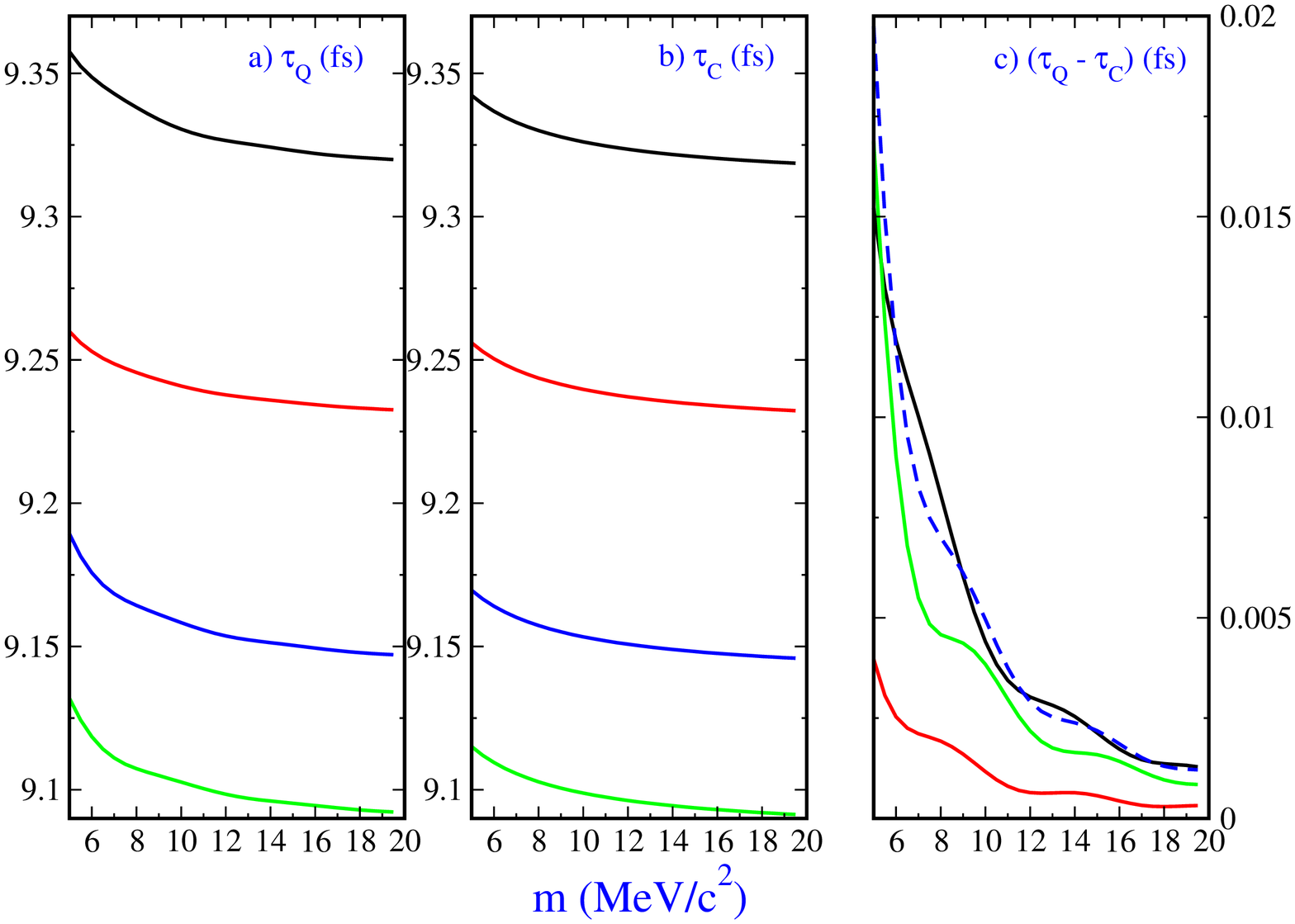}
\caption{(Color online) a) $\tau_{Q}$, b) $\tau_{C}$ and c)
$\tau_{Q} - \tau_{C}$ versus $m$ for $a=8${\AA} and $\sigma_0 =
2${\AA}; for $\alpha = -5$ (black curve), $\alpha = 0$ (red
curve), $\alpha = 2$ (green curve) and $\alpha = 5$ (blue curve).}
\vspace*{0.7cm} \label{fig: meanarrival_nonGaus}
\end{center}
\end{figure}

In Figures \ref{fig: rho_x} and \ref{fig: cur_t} we display the
probability density and probability current density for a Gaussian
wave packet for different values of mass. This density current is
positive for all times at detector location $X$. These figures
show that quantum results computed by Eqs. (\ref{eq:
qm_wave_nonGaussian}) and (\ref{eq: QM_transmitted_wave}) approach
as mass increases. For our working parameters, coincidence appears
as $m \geq 5$MeV$/$c$^2$. Further computations show that for a
given mass and barrier width $a$ these results are closer as
$\sigma_0$ increases. Moreover, for a given mass and packet width
$\sigma_0$, results become more similar as $a$ decreases. We have
also checked that (\ref{eq: qm_wave_nonGaussian}) and (\ref{eq:
QM_transmitted_wave}) give the same result as $m \geq
5$MeV$/$c$^2$, $a \leq 8$ {\AA} and $\sigma_0 \geq 2${\AA}. By
choosing the parameters in this range we compute quantum mean
arrival time $\tau_{Q}$ by means of the wave function (\ref{eq:
qm_wave_nonGaussian}). Finally, from Figure \ref{fig: meanarrival_nonGaus} one
clearly sees that for large masses and different values of
$|\alpha|$, the difference between classical and quantum mean
arrival times increase. These differences decrease with mass
and for a given value of $\alpha$. It is also apparent that
$\tau_{C} < \tau_{Q}$, which is also a result of
\cite{HoPaBa-JPA-2009} for the propagation of a
non-minimum-uncertainty Gaussian wave packet in the presence of a
linear gravitational field. However, this problem has been
computed by a different scheme, that is, the Liouville equation
instead of the classical wave equation.

%********************************************************
%********************************************************
% Figure 4
\begin{figure}
\centering
\includegraphics[width=10cm,angle=-90]{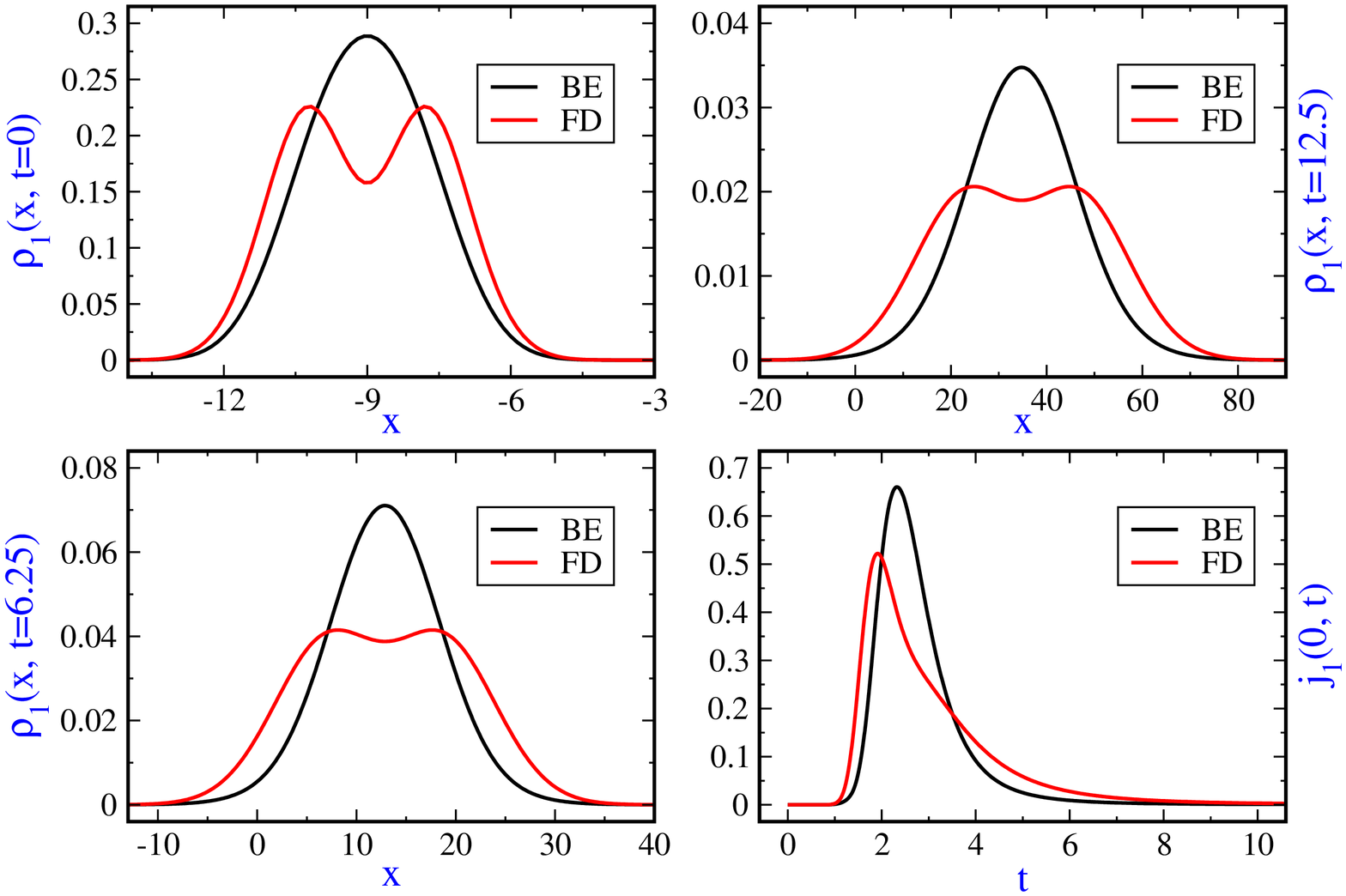}
\caption{(Color online) one-body probability density versus space
coordinate $x$ at different times and one-body probability current
density versus time at detector location $x_d = 0$. To produce this
figure we have used $m = 0.5$, $\hbar = 1$, $k_a = 2/\sigma_0$, $k_b
= 1.5/\sigma_0$, $x_{ca}= -10 \sigma_0$ $x_{cb}= -8 \sigma_0$
,$\sigma_{0a} = \sigma_{0b} = \sigma_0 $ and $\sigma_0=1$. }
\vspace*{0.5cm} \label{fig: rhosp_jsp}
\end{figure}

%
%********************************************************
%********************************************************
%
In figure \ref{fig: rhosp_jsp} we have depicted
one-body densities for a two-body system composed of two identical
particles. As this figure shows distributions of FD statistics are
wider than that of BE ones. One-body probability current
density at detector location $x_d=0$ shows arrival time distribution
in this point. For our parameters arrival of fermions at detector
location takes place sooner than bosons.

%********************************************************
%********************************************************
% Figure 5

%**********************************************************
%**********************************************************

%=================================================================================================
% Section  Section  Section  Section  Section  Section  Section  Section  Section  Section  Section
%\section{Summary and discussion}
%
%
Summarizing, in this work, quantum and classical correspondence are
studied based on the evolution of a non-Gaussian wave packet in
configuration space under the presence of a rectangular potential
barrier. Mean arrival times, at a given detector location, are
analyzed classically and quantum mechanically versus different
values of parameters of the initial wave packet and the barrier. In
particular, we have observed that (i) quantum mean arrival times are
larger than the classical ones, (ii) by increasing mass or width of
the initial wave packet classical and quantum results approach,
(iii) even though classical and quantum mean arrival times do not
have regular behavior with the non-Gaussian parameter $\alpha$,
their difference increases with $|\alpha|$ and (iv) in the range of
our parameters even though the transmitted wave packet looks like
the free evolved one,  they are displaced relative to each other due
to the first derivative of the phase of the transmission probability
in the peak momentum $p_0$. As it is widely discussed by Holland
\cite{Ho_book_1993_1} even though classical and quantum behaviors
are approaching to the same limit, it cannot be claimed that one has
deduced classical mechanics from the quantum theory in the
conventional language, because the former is a deterministic theory
of motion while the later is a statistical theory of
observation. Thus, a physical postulate (similar to the one is
arranged in the causal interpretation) must be added to quantum
mechanics. In agreement with one's intuition there is
no classical wave equation for many-body systems composed of {\it
identical} particles.
%
%{\color{red} At the end we state that the formalism presented in the
%manuscript possibly can also be applied fruitfully to hybrid systems,
%instead of systems which are either classical or quantum mechanical.
%}
At the end we state that it should be constructive
to use other approaches for computing the quantum arrival time distribution
in our example, and then compare the results of these approaches with those of
the present paper. 
\\
\\

{\bf{Acknowledgment}}

Support from the COST Action MP 1006 is acknowledged.

%=================================================================================================
% Section  Section  Section  Section  Section  Section  Section  Section  Section  Section  Section
\appendix
\section{} \label{sec: app}

In this appendix some details on the non-Gaussian wave packet are
provided. This corresponding initial wave packet is built from the
amplitude function (\ref{eq: initial_amplitude}) and is actually a
superposition of three Gaussian wave packets with the same center
$x_c$ but with different kick wave vectors $k_0$, $k_+$ and $k_-$.
The expectation value of the position operator and the uncertainty
in position are respectively,
\begin{eqnarray}
\langle x \rangle &=& x_c + \sigma_0 \frac{\pi
e^{3\pi^2/32}\alpha}{\alpha^2(e^{\pi^2/8}-1)+2e^{\pi^2/8}}~, \label{eq: xave_nonG}\\
\sigma_x &=& \sqrt{\langle x^2 \rangle - \langle x \rangle^2}
\nonumber \\
&=& \sigma_0 \sqrt{1+ \frac{\pi^2 \alpha^2
[\alpha^2(e^{\pi^2/8}-1)+2e^{\pi^2/8}-4e^{3\pi^2/16}]}
{4[\alpha^2(e^{\pi^2/8}-1)+2e^{\pi^2/8}]^2}}~. \label{eq: sigmax}
\end{eqnarray}
The Fourier transform of the initial wave packet is
\begin{eqnarray} \label{eq: Fourier_nonGaussian}
\phi(k) &=& \left(\frac{2 \sigma_0^2}{\pi}\right)^{1/4} \frac{
e^{-\sigma_0^2(k-k_0)^2 } } {\sqrt{1 +
\frac{\alpha^2}{2}(1- e^{-\pi^2/8})}} \nonumber \\
&\times& \left[ 1 - \frac{i~\alpha}{2} e^{-\pi^2/16} \left(
e^{\frac{\pi \sigma_0}{2} (k-k_0)} - e^{-\frac{\pi \sigma_0}{2}
(k-k_0)} \right) \right]~,
%&=&
%\left(\frac{2 \sigma_0^2}{\pi}\right)^{1/4}
%\frac{ e^{ - i k x_c} } {\sqrt{1 + \frac{\alpha^2}{2}(1- e^{-\pi^2/8})}}
%\left[ e^{ -\sigma_0^2(k-k_0)^2 } - \frac{i~\alpha}{2}
%\left(
%e^{ -\sigma_0^2(k-k_0- \frac{\pi}{4\sigma_0})^2 } - e^{ -\sigma_0^2(k-k_0 + \frac{\pi}{4\sigma_0})^2 }
%\right)
%\right]~,
\end{eqnarray}
and the expectation value of the momentum operator and the
uncertainty in momentum are respectively (in terms of wave
vectors),
\begin{eqnarray}
\langle k \rangle &=& k_0 ~, \\
\sigma_k &=& \sqrt{\langle k^2 \rangle - \langle k \rangle^2} =
\frac{1}{2\sigma_0} \sqrt{ 1 + \frac{ \alpha^2 \pi^2 }
{8 + 4 \alpha^2 ( 1 - e^{-\pi^2/8} ) } } \label{eq: sigmak}
\end{eqnarray}
\begin{figure}
\begin{center}
\includegraphics[width=7cm,angle=-90]{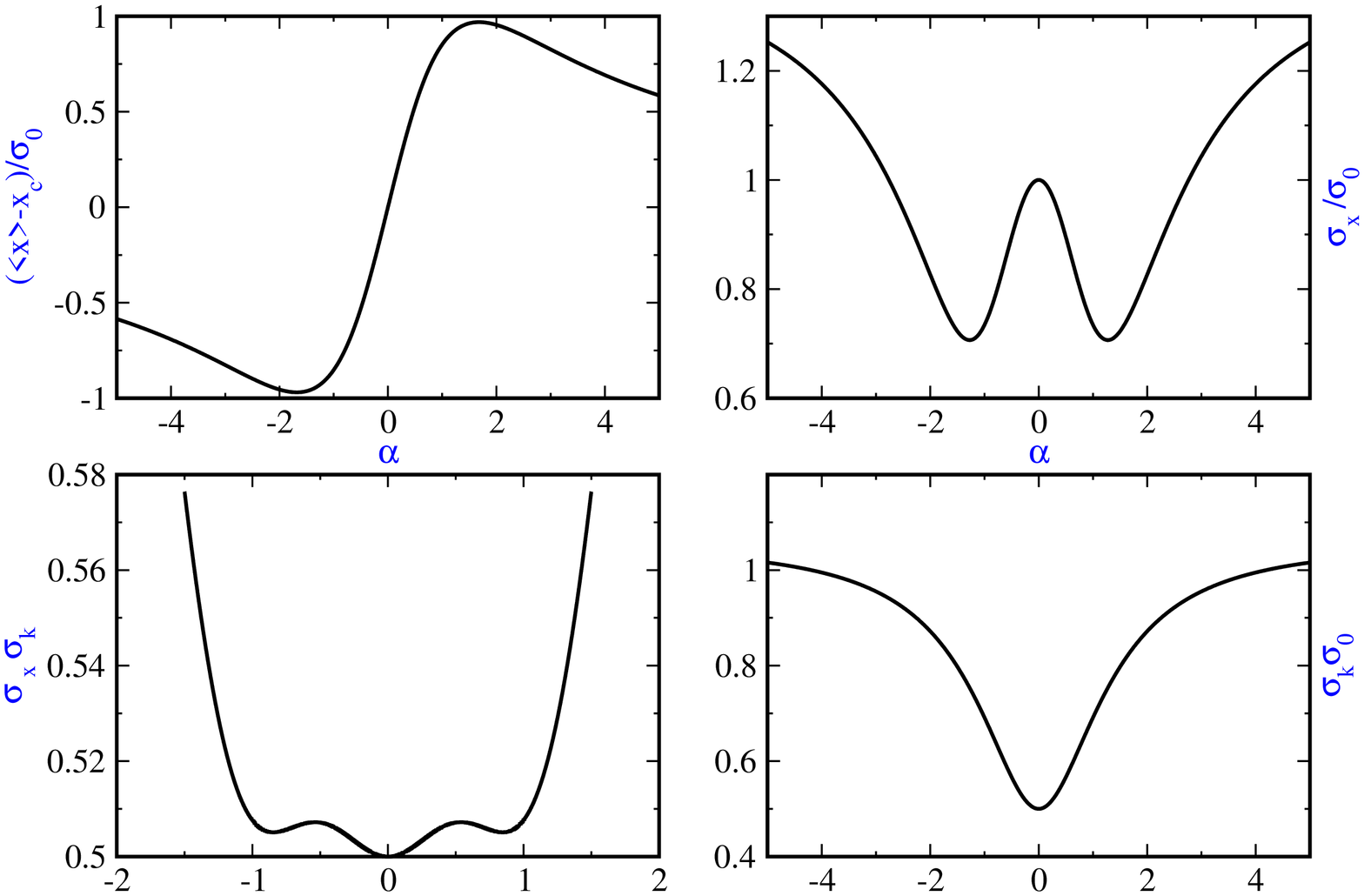}
\caption{(Color online) $\langle x \rangle$, $\sigma_x$,
$\sigma_k$ and $\sigma_x \sigma_k$ versus $\alpha$ for the initial
non-Gaussian wave packet.} \vspace*{0.7cm} \label{fig:nonGaussian}
\end{center}
\end{figure}
One clearly sees that $\sigma_x$, $|\phi(k)|^2$ and $\sigma_k$ are
even functions of $\alpha$. The function $\rho(k) = |\phi(k)|^2$
is symmetric around central wave number $k_0$; and this point is a
global maximum for $ - \alpha_0 < \alpha < \alpha_0$ while is a
local minimum for $ \alpha < - \alpha_0 $ or $ \alpha_0 < \alpha$,
where $\alpha_0 = 2 \sqrt{2} \exp{[\pi^2/16]}/\pi = 1.66836$. In
Figure \ref{fig:nonGaussian} the expectation value of position,
uncertainty in position, uncertainty in momentum and the product
of uncertainties versus $\alpha$ are plotted. Finally, the
propagation of the above non-Gaussian packet in free space is
given by
\begin{eqnarray}  \label{eq: free_nonGaussian}
\psi(x, t) &=&
 \frac{1}{\left( 2\pi\sigma_0^2 \left[ 1 + i\frac{\hbar t }{2m\sigma_0^2} \right]^2 \right) ^{1/4} }
 \frac{1}{\sqrt{ 1+ \frac{\alpha^2}{2} (1-e^{ -\pi^2 /8}) } }
 \nonumber \\
&\times& \left[ 1 + \alpha~e^{\frac{\pi^2}{16} \left( \frac{1}{ 1
+ i\frac{\hbar t }{2m\sigma_0^2} }-1 \right) } \sin \left( \pi
\frac{(x-x_c)-u t }{4 \sigma_0 \left[ 1 + i\frac{\hbar t
}{2m\sigma_0^2} \right]} \right) \right]
\nonumber \\
&\times &
\exp{ \left\{ i k_0 \left[x -\frac{u}{2}t \right] \right\} }  ~
\exp{ \left\{ -\frac{[(x-x_c)-u t ]^2}{ 4 \sigma_0^2 \left[ 1 + i\frac{\hbar t }{2m\sigma_0^2} \right] } \right\} }~.
\end{eqnarray}
%

%********************************************************


\begin{thebibliography} {99}

%
\bibitem{Ho_book_1993_1}
P. R. Holland 1993, {\it The Quantum Theory of Motion} (Cambridge:
Cambridge University Press), chapter 6 and references therein
%
\bibitem{Ba-book-1998}
L. E. Ballentine 1998 {\it Quantum Mechanics: A Modern
Development} (Singapore: World Scientific), chapter 14
%
\bibitem{Home-book-1997}
D. Home 1997 {\it Conceptual Foundations of Quantum Physics: An
Overview from Modern Perspectives} (New York: Plenum), chapter 3 ; \\
D. D\"{u}rr, S. Goldstein and N. Zanghi 2013 {\it Quantum Physics
Without Quantum Philosophy} (Berlin Heidelberg: Springer-Verlag)
chapter 5
%
\bibitem{antonio2013}
A. B. Nassar and S. Miret-Art\'es, Phys. Rev. Lett. {\bf 111},
150401 (2013)

\bibitem{FeLeSa-book-1965}
R. P. Feynman, R. B. Leighton and M. Sands 1965 {\it Lectures on Physics}
vol. III, Addison--Wesley, Reading, MA
%
\bibitem{St-RMP-1966}
F. Strocchi, Rev. Mod. Phys. {\bf 38} (1966) 36
%
\bibitem{He-PRD-1985}
A. Heslot, Phys. Rev. D {\bf 31} (1985) 1341
%
\bibitem{Elz-PRA-2012}
H-T. Elze {\it Phys. Rev. A} {\bf 85} (2012) 052109
%
\bibitem{FrLaEl-arxiv-2014}
L. Fratinoa, A. Lampob and H.-T. Elze (2014) arXiv:1408.1008
%
\bibitem{LaFrEl-arxiv-2014}
A. Lampo, L. Fratino and H.-T. Elze (2014) arXiv:1410.4472
%
\bibitem{HoPaBa-JPA-2009}
D. Home, A. K. Pan and A. Banerjee {\it J. Phys. A} {\bf 42} (2009) 165302
%
\bibitem{Ri-JPA-2013}
N. Riahi {\it J. Phys. A} {\bf 46} (2013) 208001
%
\bibitem{HoPaBa-JPA-2013}
D. Home, A. K. Pan and A. Banerjee {\it J. Phys. A} {\bf 46} (2013) 208002
%
%
\bibitem{Ro-AJP-1964}
N. Rosen, {\it Am. J. Phys.} {\bf 32} (1964) 377
%
\bibitem{Ro-AJP-1965}
N. Rosen, {\it Am. J. Phys.} {\bf 33} (1965) 146
%
\bibitem{CoDiLa_book_1977}
C. Cohen-Tannoudji, B. Diu, F. Lal\"{o}e, 1977 {\it Quantum
Mechanics} (Paris: Wiley); \\ A.E. Bernardini, {\it Ann. Phys.}
{\bf 324} (2009) 1303
%
\bibitem{Wi-PR-2006}
H. G. Winful, {\it Phys. Rep.} {\bf 436} (2006) 1
%
\bibitem{Ho_book_1993_2}
P. R. Holland 1993, {\it The Quantum Theory of Motion} (Cambridge:
Cambridge University Press) pp 55-61
%
%
\bibitem{MuLe-PR-2000}
J. G. Muga and C. R. Leavens {\it Phys. Rep.} {\bf 338} (2000) 353
%
\bibitem{Ho-PRA-1999}
P. R. Holland {\it Phys. Rev. A} {\bf 60} (1999) 4326
%
%
\bibitem{StBaNeWe-PLA-2004}
W. Struyve, W. De Baere, J. De Neve and S. De weird  {\it Phys. Lett. A} {\bf 322} (2004) 24
%
%
\bibitem{ChHoMaMoMoSi-CQG-2012}
P. Chowdhury, D. Home, A. S. Majumdar, S. V. Mousavi,
M. R. Mozaffari and S. Sinha {\it Class. Quantum Grav.} {\bf 29} (2012) 025010
%
%
\bibitem{Zelev-book-2011}
V. Zelevinsky 2011, {\it Quantum Physics: From Time-Dependent Dynamics to Many-Body Physics and Quantum Chaos} (Weinheim:
Wiley-VCH), page 393
%
\end{thebibliography}
\end{document}